\documentstyle[prd,aps]{revtex}

\def\case#1#2{\hbox{$\frac{#1}{#2}$}}
\def\slantfrac#1#2{\hbox{$\,^#1\!/_#2$}}
\def\ion#1#2{#1$\;${\small\rm\@Roman{#2}}\relax}

\begin{document}
\draft

\title{Curvature pressure: Sufficient for a static and stable cosmology;\\
 important for solar neutrino production and black hole formation}

\author{David F. Crawford}
\address{School of Physics, A28, University of Sydney, N.S.W. 2006, Australia\\
d.crawford\@physics.usyd.edu.au}
\date{\today}
\maketitle
\begin{abstract}

A hypothesis is presented that electromagnetic forces that prevent ions from following
geodesics results in a curvature pressure that is very important in astrophysics. 
It may partly explain the solar neutrino deficiency and it may be the engine that 
drives astrophysical jets. 
However the most important consequence is that it leads to a static and stable
cosmology using general relativity without a cosmological constant. 
Combined with an earlier hypothesis of a gravitational interaction of photons and 
particles with curved spacetime a static cosmology is developed that predicts a Hubble
constant of $H=60.2\,\mbox{km}\mbox{s}^{-1}\mbox{Mpc}^{-1}$and a microwave background 
radiation with a temperature of 3.0\,K. 
The background X-ray radiation is explained and observations of the 
quasar luminosity function and the angular distribution of radio sources have a better
fit with this cosmology than they do with standard big-bang models.  
Observations that require dark matter in big-bang cosmologies are explained as being
due to redshifts due to gravitational interaction in  intervening clouds and therefore
no dark matter is required.  
The result is a static and stable cosmological model that agrees with most of the current
observations.
\end{abstract}

\pacs{98.80, 26.65, 95.35, 98.70.V}

\section{Introduction}
In previous papers \cite{Crawford87a,Crawford91} it was argued  that there is a gravitational 
interaction  such that photons and particles  lose energy as they pass through a gas. 
The energy loss for photons' results in a redshift that could produce the Hubble redshift.
A `tired light' mechanism to explain the Hubble redshift entails a static cosmology 
which was developed  within a Newtonian context \cite{Crawford93}. This paper  presents
a further development of a static, stable cosmology within the framework of general
relativity. It is based on the observation  that in plasmas electromagnetic forces 
completely dominate the particle motions  so that they do not travel along geodesics. 
The new hypothesis is that there is  a reaction back on the material that generates 
curved spacetime. Where curved spacetime  is due to a plasma the reaction  is seen
as a (curvature) pressure within the plasma that depends on its density and temperature
and acts to prevent compression.  

The curvature pressure is investigated in a static cosmological model and for plasmas that 
occur in the center of the sun and around compact objects. It is shown that the effect of 
curvature pressure will  decrease the central solar temperature by an amount that may be  
sufficient to explain the observed deficiency of solar neutrinos.  
Since curvature pressure acts to oppose contraction and since it increases with 
temperature it is unlikely that black holes could form from hot plasmas. However
it remains  possible to form black holes from cold material. More significantly 
curvature pressure is very important in accretion disks around compact objects and
may provide the engine that drives astrophysical jets. 

Since the big-bang cosmological model in all its ramifications is so well entrenched,  
to be taken seriously,  any alternative model must at least be able to explain 
the major cosmological observations.
It is argued that using the Friedmann equations the introduction of curvature pressure
leads to a static and stable cosmological model. One of the predictions of this model 
is that there is a background X-ray radiation and an analysis of the background 
observations done in a previous paper are used to determine the average density of 
the universe. 
Because of its essential importance to this static cosmology and because the earlier
results did not include the effects of curvature pressure the hypothesis of a 
gravitational interaction is revisited. 
The result is a prediction  of the Hubble constant and the microwave background
radiation. It is  shown how the observations that lead to the occurrence of dark 
matter in the big-bang cosmology are readily explained without dark matter.
Next previous work on the luminosity function of quasars and the angular sizes 
of radio sources is discussed to show that the observations can be fitted without 
evolution. 
The theme of evolution, or lack of it, is continued with examination of observations
on quasar absorption lines, a  microwave background temperature at high redshif
t and type 1a supernovae light curves. 
Finally the topics of nuclear abundance, entropy, and Olber's paradox are briefly 
covered.  

\section{Theoretical background}
A theme that is common to the development of both curvature pressure and the gravitational 
interaction is that in four-dimensional-space the effects of centripetal acceleration 
are essentially  the same as they are in three-dimensional-space.
Consider two meridians of longitude at the equator  with a perpendicular separation of 
$h$, then  as we move along the longitudes this separation obeys  the differential
equation   $h''=-h/r^2$ where the primes denote differentiation with 
respect to the path length and  $r$ is  the radius of the earth. 
In addition the particle has a centripetal acceleration of   $v^2/r$ where  $r$  can
be determined from the behavior of  $h''$. 
In four-dimensional-space the longitudes become a geodesic bundle and the separation 
becomes a cross-sectional area, $A$, where  $A''=-A/r^2$.
Again the particle has a centripetal acceleration of  $v^2/r$ where now  $r$ is the 
radius of the hyper-sphere. Although the particle as we know it is  confined to three 
dimensions there is a centripetal acceleration due to curvature in the fourth  
dimension that could have significant effects.  
Another fundamental topic considered is the nature  of gravitational force.
It is critical to the development of curvature pressure that gravitation  produces 
accelerations and not forces.

The gravitational interaction theory explicitly requires that photons and particles 
are described  by localized wave packets. 
The wave equations that describe their motion in flat spacetime are  carried over to 
curved spacetime in which the rays coincide with geodesics. 
In particular with the  focusing theorem \cite{Misner73}  there is 
an actual focusing of the wave  packet in that its cross-sectional area decreases as 
the particle (photon)  travels along its trajectory.  
In this and previous papers \cite{Crawford87a,Crawford91} it is argued that the result is 
a gravitational  interaction in which the particle  loses energy.

\section{The theoretical model for curvature pressure}
In a plasma there are strong, long-range electromagnetic forces that completely dominate 
accelerations due to gravitational curvature. 
The result is that, especially for electrons, the particles do not travel along geodesics. 
If we stand on the surface of the earth our natural geodesic  is one of free fall 
but the contact forces of the ground balance the gravitational acceleration with  
the consequence that there is a reaction force back on the ground. The result of 
stopping our  geodesic motion is to produce a force that compresses the ground.
The major hypothesis of this paper is that there is a similar reaction force in four-dimensional 
spacetime. This force acts back on the plasma (that produces the curved spacetime) because its 
particles do not follow geodesics. Thus the plasma appears in two roles. 
The first produces the curved spacetime and in the second the failure of its particles 
to follow geodesics causes a reaction back on itself acting in the first role. 
It is the long-range electromagnetic forces that are important, not particle collisions. 
For example in a gas without long-range forces and assuming that the time spent 
during collisions is negligible  the particles will still travel along geodesics between collisions and there is no reaction. 
Given that there are long range forces that dominate the particle trajectories 
there is a  reaction force that appears as a pressure, the curvature pressure. Just as we do not need to know the details of the contact
forces when we are stationary on the ground in order to calculate the reaction force, we can compute the reaction force in a plasma from
the gravitational accelerations. 

For the cosmological model consider  the plasma to occupy the surface of a four-dimensional hypersphere.  
It is easier to imagine if one of the normal dimensions is suppressed then it will 
appear as the two-dimensional surface of a three-dimensional sphere. 
The nature of this pressure can then be understood by analyzing this reduced model 
another way of describing the effects of the centripetal accelerations of the particles. 
By symmetry the gravitational attraction on one particle due to the rest is equivalent 
to having the total mass at the center of the sphere.
To start let the shell contain identical particles all with the same velocity, and let 
this sphere have  a radius r,  then the radial acceleration of a particle with 
velocity v is $v^2/r$. 
At equilibrium the radial accelerations are balanced by the mutual gravitational attraction.
Now for a small change in radius, $dr$, without any change in the particle velocities 
and going from one equilibrium position to another  we can equate the work done by the 
curvature pressure to the work done by the force required to overcome the centripetal 
acceleration to get
\[
p_cdA=-{{Mv^2} \over {r}}dr,
\]
where M is the total mass, but for a two-dimensional area  $dA/dr=2A/r$ therefore  
$p_c=-Mv^2/2Ar=-\rho v^2/2$ where $\rho $ is the surface density.  
Thus the effects of the centripetal accelerations can be represented as a negative 
pressure acting within the shell. 
The next step is to generalize this result to many types of particles where each type 
has a distribution of velocities. 
 
The particles are constrained  to stay in the shell by a dimensional constraint that 
is not a force.  
The experiments of E\"{o}tv\"{o}s and others (Roll et al. 1964 and Braginski and Panov 1971) 
show that the Newtonian passive gravitational mass is identical to the inertial mass to 
about one part in  $10^{12}$. 
The logical conclusion is that Newtonian gravitation produces an acceleration and not a force. 
The mass is only introduced for consistency with Newton's second law of motion. 
The concept of gravitation as an acceleration and not a force is even stronger in general 
relativity. 
Here the geodesics are the same for all particles independent of their mass and 
gravitational motion does not use the concept of force. 
Clearly for a single type of particle the averaging over velocities is straightforward 
so that the curvature pressure is $p_c=-\rho \overline {v^2}/2$. 
The averaging over particles with different masses is more ambiguous. 
Traditionally we would weight the squared velocities by their masses; that is we 
compute the average energy. 
However since the constraint that holds the particles within the two-dimensional shell 
is not due to forces and since gravitation  produces accelerations and not forces the 
appropriate average is over their accelerations.  
The result for our simple Newtonian model is  
\[
p_c=-\case{1}{2}\rho \sum\limits_i {\overline {v_i^2}},
\] 
where  the density is defined as   $\rho =\sum\limits_i {n_i}m_i$ and $n_i$ is the 
number density of the i'th type of particle. 
This simple Newtonian  model gives a guide to what the curvature pressure would be 
for a more general model in a homogeneous isotropic three-dimensional gas that forms 
the surface of a four-dimensional hyper-sphere. 
The dimensional change requires that we replace  $dA/dr$  by $dV/dr=V/3r$, and  then 
including the relativistic corrections (a factor of  $\gamma ^2$) needed to transform 
the accelerations from the particle's reference system to a common system where the 
average velocity is zero, we get
\begin{eqnarray}
\label{e3}
p_c& = &-{{\rho}  \over {3}}\sum\limits_i {n_i}\overline {\gamma _i^2v_i^2}\nonumber\\
 \nonumber
& =& -{{\rho c^2} \over {3}}\sum\limits_i {n_i\left( {\overline {\gamma _i^2}-1}
 \right)}\nonumber\\
& =&  -{{\rho c^2} \over {3}}{\left( \overline {\gamma^2}-1 \right)},
\end{eqnarray}
where  the Lorentz factor $\gamma ^2=1/\sqrt {1-v^2/c^2}$. 
Note that although the equation for curvature pressure does not explicitly include 
the spacetime curvature the derivation requires that it is not zero. 
Because this equation was only obtained by a plausibility argument we hypothesise that 
the curvature pressure in the cosmological model is given by 
equation (\ref{e3}).

Since  the particles may have relativistic velocities, and assuming thermodynamic 
equilibrium,  the   ($\overline {\gamma ^2}-1$)  factor can be evaluated using the 
J\"{u}ttner distribution.  
For a gas at temperature  $T$ and particles with  mass  $m$  de Groot, 
Leeuwen \& van Weert \cite{Groot80}  show that 
\begin{equation}
\label{e4}
\gamma ^2(\alpha )=3\alpha K_3(1/\alpha)/K_2(1/\alpha)+1,
\end{equation} 
where $\alpha = kT/mc^2$ and  $K_n(1/\alpha)$ are the modified Bessel functions of 
the second kind \cite{Abramowitz72}. 
For small  $\alpha $ this has the approximation
\begin{equation}
\label{e5}
\gamma ^2(\alpha )=1+3\alpha +{\case{15}{2}}\alpha ^2+{\case{45}{8}}\alpha ^3+\ldots .
\end{equation}
Note for a Maxwellian distribution the first three terms are exact so that the extra 
terms are corrections required for the J\"{u}ttner distribution. 
For non-relativistic velocities equation (\ref{e5}) can be used and equation 
(\ref{e3}) becomes
\[
p_c=-\frac{1}{n}\sum\limits_{i=1}^N {\left( {{{n_i} \over {m_i}}}\right)}\overline mkT,
\]
where  $n_i$ is the number density for the i'th type of particle and   
$\overline m=\sum\limits_{i=1}^N {n_im_i}/n$ is the mean particle mass. 
Except for the 
inverse mass weighting  and the sign this is identical to the expression for 
the thermodynamic pressure.
 
\section{Solar interior  and local plasma concentrations}
The equation for curvature pressure derived above for the cosmological model cannot 
be used in other situations with different metrics. 
The key to understanding the application of curvature pressure in other metrics such 
as the Schwartzschild metric used for stellar interiors is to consider the case where 
the overall curvature is small and superposition  may be assumed. 
Since the free fall acceleration of a particle is independent of its mass there is no 
curvature pressure associated  with external gravitational fields provided they have 
scale lengths much greater than the typical ion separation.  
Any curvature pressure is due to local curvature of the metric produced by the local 
density.  
This arises because although the electrons and ions have in general different centripetal 
accelerations these are completely dominated by accelerations due to the electromagnetic 
forces. 
Let the gravitational potential be   $\Phi $, then the self-gravitational energy density 
is $\rho\Phi $. 
Now it was argued above that the curvature pressure is proportional to the energy density 
(it has the same units) but with an averaging over accelerations rather than forces that 
results in replacing   $\rho$ by  $\left( {\overline {\gamma ^2}-1} \right)\rho $. 
Consequently we take the curvature pressure in a plasma due to its own density as
\begin{equation}
\label{e7}
p_c=\case{1}{3}\left( {\overline {\gamma ^2}-1} \right)\rho \Phi 
\end{equation}  
Note that the derivation is essentially one based on 
dimensional analysis and therefore the numerical factor of \slantfrac{1}{3} may need modification. 
It was used in part for consistency with the cosmological curvature pressure and in part 
because it makes the application of equation (\ref{e7}) to a low temperature gas with a 
single type of particle have the simple expression   $p_c=p_T\Phi /c^2$ where   $p_T$ 
is the thermodynamic pressure.
From potential theory we get for the curvature pressure of a plasma at the point $r_0$ 
the expression
\begin{equation}
\label{e8}
p_c\left( {r_0} \right)=
\case{1}{3}G\rho \left( {r_0} \right)\left( {\overline {\gamma 
^2\left( {r_0} \right)}-1} \right)
\int \frac{\rho \left( {r-r_0} \right)}{\left| {r-r_0} \right|} dV
\end{equation}.
Equation (\ref{e8}) can be simplified for non-relativistic velocities by using the 
approximation (equation \ref{e5}) to get 
 \begin{equation}
\label{e9}
p_c={{G\rho \left( {r_0} \right)kT} \over {c^2}}\left( {\sum\limits_{i=1}^N 
{{{n_i} \over {nm_i}}}} \right)
\int \frac{\rho \left( {r-r_0} \right)}{\left| {r-r_0} \right|} dV
\end{equation} 
where n is the total number density. 

The curvature pressure adds to the thermodynamic pressure (and radiation pressure) to 
support the solar atmosphere 
against its own gravitational attraction. 
That is for the same gravitational attraction the required thermodynamic pressure, and 
hence the temperature, will be reduced by curvature pressure. 
Applying equation (\ref{e9}) to the sun and using pressures, temperatures, and abundance 
ratios given by Bahcall  \cite{Bahcall89}, it was found that the curvature pressure at the center 
of the sun is  $2.8\times 10^{14}\,\mbox{Pa}$ compared to the thermodynamic pressure 
of   $2.3\times 10^{16}\,\mbox{Pa}$. 
Since the temperature is directly proportional to the thermodynamic pressure this 
implies that the temperature at the center of the sun is  reduced by 1.2\%.
Bahcall  \cite{Bahcall89} shows that the $^8$B neutrino flux is very  sensitive to the 
temperatures at the center of the sun with a flux rate that is proportional to the 
eighteenth power of the temperature. 
Thus this temperature change would decrease the neutrino flux to 80\% of  that from 
the standard model. 
Although the observed ratio of   $2.55/9.5=27\% $ \cite{Bahcall97}  is much smaller 
the effect of the pressure curvature is clearly significant and large enough to 
warrant a more sophisticated computation.

\section{Black holes and astrophysical jets}
One of the major mysteries in current astrophysics is the occurrence of powerful 
relativistic jets on both stellar and galactic scales. 
Could curvature pressure produce the force that drives these jets? It has the basic 
requirements of being able to provide the energy and it is clearly an important 
factor in the accretion disk around any compact object.  
Another important application of the curvature pressure is in the formation of black holes. 
By the time densities of neutron stars are reached the strong nuclear force provides 
the force that prevents the particles traveling along geodesics. 
Basically the curvature pressure will prevent the formation of hot black holes. 
There is still the possibility of cold black holes if material can be accreted without 
getting hot enough for the curvature pressure to be significant. 
Even if the curvature pressure can prevent the formation of a black hole there would 
still be a compact object that from the outside would appear almost identical to a black hole. 
Thus the model is consistent with the observations of very massive compact objects 
\cite{Rees97,Begelman96,Kormendy95}.

\section{The cosmological model with curvature pressure}
The main application of curvature pressure is to a  cosmological model  for a 
homogeneous and isotropic distribution of a fully ionized gas. 
Based on the theory of general relativity and using the Robertson-Walker metric 
the Friedmann equations \cite{Weinberg72} are
\begin{eqnarray*}
-\ddot R & =& {{4\pi G} \over {c^2}}\left( {\rho c^2+3p}\right)R\\
 R\ddot R+2\dot R^2 & = & {{4\pi G} \over {c^2}}\left( {\rho c^2-p} 
\right)R^2 -2kc^2,
\end{eqnarray*}  
where R is the radius $\rho $ is the proper density,  $p$ is the pressure, $G$ is 
the Newtonian gravitational constant, and  $c$ is the velocity of light. 
The constant  $k$ is one for a closed universe, minus one for an open universe and 
zero for a universe with zero curvature.  
Working to order  $m_e/m_H$ the thermodynamic pressure can be neglected but not the 
curvature pressure.  
The equations including the curvature pressure (equation \ref{e3}) are
\begin{eqnarray*}
-\ddot R & = & 4\pi G\rho R\{ 1-\left( {\overline {\gamma ^2}-1} \right)\} \\
R\ddot R+2\dot R^2 & =&  4\pi G\rho R^2\{ 1+\case{1}{3}\left( {\overline {\gamma^2}-1} \right)\} \\
& & - 2kc^2, 
\end{eqnarray*}
where  $\overline {\gamma ^2}$ is the average over all velocities and particle types.  
Clearly 
 $\ddot R$ is zero if   $\overline {\gamma ^2}=2$ and equation (\ref{e4})  can be 
solved for a hydrogen plasma to get   $\alpha _e=kT_0/m_ec^2=0.335$ or   
$T_0=1.99\times 10^9$K. 
Thus with thermal equilibrium the second derivative of  $R$ is zero if the plasma 
has this temperature.  
This temperature is based on a model in which the plasma is homogeneous, but the 
occurrence of galaxies and clusters of galaxies show that it is far from homogeneous. 
In order to investigate the effects of inhomogeneity consider a simple and quite 
arbitrary model where the plasma is clumped with the probability of a clump having 
the density n is given by the exponential distribution  $\exp \left( {-n/n_0} \right)/n_0$,  
where  $n_0$n  is the average density. 
Assuming pressure equilibrium so that  $T_e=T_0n_0/n$ then for  $\overline {\gamma ^2}=2$ 
we find that  the average temperature $T=1.1\times 10^9$K thus showing that the effect 
of inhomogeneity  could reduce the observed temperature
by a  factor of  order two. 

Since the right hand side of the second Friedmann equation is positive then the 
curvature constant $k$ must be greater or equal to zero. 
The only useful static solution requires that   $k=1$ and with  $\dot R=\ddot R=0$ 
the result for the radius of the universe is given by
 \begin{equation}
\label{e14}
{{1} \over {R_0^2}}={{8\pi G\rho _0} \over {3c^2}}.
\end{equation}
Thus the model is a static cosmology with positive curvature. 
Although the geometry is the 
same as the original Einstein static model  this cosmology  differs in that it does 
not require a cosmological constant. 
Furthermore it is stable. Consider a perturbation,  $\Delta R$,  about the equilibrium 
position then the perturbation equation is
 \[
\Delta \ddot R={{c^2} \over {8\pi R_0}}\left( {{{d\overline {\gamma 
^2}} \over {dR}}} \right)\Delta R,
\]
and since for any realistic equation of state the average velocity  (temperature) 
will decrease as R increases the right hand side is negative showing that the result 
of a perturbation is an oscillation about the equilibrium value. 
Thus this model does not suffer from the deficiency that the static Einstein model 
has of gross instability.  
Since the volume of the three-dimensional surface of the hyper-sphere is  
$2\pi ^2R_0^3$ the radius of the universe can be written  in terms of the total mass 
of the universe, $M_0$, as
\[
R_0={{4GM_0} \over {3\pi c^2}},
\]
which differs by a factor of $2/3$ from that \cite{Crawford93} which was derived from 
a purely Newtonian model. 
For interest the values with a density of  $2.05m_H \mbox{ m}^{-3}$ (see below) are  
$R_0=2.17\times 10^{26\,}\mbox{m}\ =7.04\,\mbox{Gpc}$, and  $M_0=6.90\times 10^{53}\,\mbox{kg}\ =3.47\times 10^{23}\,\mbox{M}_{\mbox{sun}}$.

\section{Background X-ray radiation }
If this cosmological model is correct there should be a very hot plasma between the 
galaxies and in particular between galactic clusters. 
This plasma should produce a diffuse background X-ray radiation and indeed such 
radiation  is observed.  
Attempts to explain the X-rays by bremsstrahlung radiation within the standard 
model have not been very successful \cite{Fabian92}, mainly because it must 
have originated at earlier epochs when the density was considerably larger 
than present. 
The hard X-rays could come from disctete sources but if it did there are problems
with the spectral smoothness and strong evolution is required to achieve the
observed flux density \cite{Fabian92}.
However there is an excellent fit to the data in a static cosmology 
\cite{Crawford87b,Crawford93} 
for X-ray energies between 5\,KeV and 200\,KeV. 
Using universal abundances \cite{Allen76} the analysis showed a temperature of 
$1.11\times 10^9$\,K and a density of  $2.05m_H \mbox{ m}^{-3}$.  
Comparison of this temperature with that predicted by the homogeneous model of
$1.99\times 10^9\,\mbox{K}$ shows that it is nearly a factor of two too small.  
A possible explanation comes from the observation that the universe is not homogeneous. 
Although there is fortuitous agreement with the simple inhomogeneous model described 
above this can only be interpreted as showing that the observations are consistent with 
an inhomogeneous model.  

One of the main arguments against the explanation that the background X-ray radiation 
comes from a hot inter-cluster plasma is that this plasma would distort the cosmic 
microwave background radiation  by the Sunyaev-Zel'dovich effect. 
This distortion is usually expressed by the dimensionless parameter $y$.  
Mather et al \cite{Mather94} have measured the spectrum of the cosmic microwave background 
radiation and conclude that  $\left| y \right|<2.5\times 10^{-5}$. 
In the big-bang cosmology most of the distortion occurs 
at earlier epochs where the predicted density and the temperature of the plasma are 
much higher than current values. 
However for any  static model we can use a constant density of 
$2.05m_H \mbox{ m}^{-3}$ in the equation \cite{Peebles93}
\[
y={{kT_e\sigma _Tn_er} \over {m_ec^2}},
\]
where  $\sigma _T$ is the Thomson cross-section and $r$  is the path length since 
the formation  of the radiation. 
For a hydrogen plasma we get  $y=2.6\times 10^{-29}r$. 
In this model (see below) the background radiation is being 
continuously replenished by energy losses from the hot electrons and the  typical 
path length for the energy lost by electrons to equal the energy of a photon at the 
peak of the spectrum is  
$3.5\times 10^{18}\,$m which results in  $y=9.1\times 10^{-11}$ well within the 
observed limits.

\section{The Hubble constant}
One  of the major requirements of any cosmological model is the necessity to explain the 
relationship found by Hubble that the redshift of extra-galactic objects depends on 
their distance.  
In earlier papers \cite{Crawford87a,Crawford91} the author suggested that there is an 
interaction of photons with curved spacetime that  produces an energy loss that 
can explain the Hubble redshift relationship.  
Because the earlier work did not include the effects of curvature pressure and because 
this interaction is central to the description of a viable static cosmology a brief  
updated description is given here. 
The principle is that a photon can be considered as a localized wave traveling along a 
geodesic bundle. 
Because of the `focusing theorem' \cite{Misner73} the cross-sectional area of  this 
bundle will decrease with time, and in applying this theorem to a photon it was argued 
that this will cause a change in the photon's properties. 
In particular angular momentum will decrease because it is proportional to a spatial 
integral over the cross-sectional area.  
The change in angular momentum can only be sustained for a time consistent with the 
Heisenberg uncertainty principle. 
The conclusion is that eventually there will the emission of two (in order to conserve 
the total angular momentum)  very low energy photons.

The second part of the argument is that the rate at which this energy loss occurs is 
proportional to the rate of change of area of the geodesic bundle.  
This rate of change of area in the absence of shear and vorticity is given by the 
equation \cite{Raychaudhuri55}, 
\[
{{1} \over {A}}{{d^2A} \over {ds^2}}=-R_{\alpha \beta }U^\alpha U^\beta ,
\]
where   $R_{\alpha \beta }$ is the Ricci tensor,   $U^\alpha $ is the four-velocity 
and, s is a suitable affine parameter. 
At any point the trajectory of the  geodesic bundle is tangential to the surface of 
a four-dimensional hyper-sphere with radius $r$.  
Then since the centripetal acceleration is  $c^2/r$ where   $r$ is defined by 
\[
{{1} \over {r^2}}={{1} \over {A}}{{d^2A} \over {ds^2}}
\] 
we can define   $\varepsilon $, the fractional rate of energy loss by 
\[
\varepsilon =c^2\sqrt {{{1} \over {A}}{{d^2A} \over {ds^2}}}.
\]
This relationship for   $\varepsilon $ is a function only of Riemann geometry and 
does not depend on any particular gravitational theory. 
However, Einstein's general relativity gives a particularly elegant evaluation. 
Direct application of the field equations with the stress-energy-momentum tensor 
$T_{\alpha \beta }$ gives
\[
\varepsilon =\sqrt {{{8\pi G} \over {c^2}}\left( {T_{\alpha \beta }U^\alpha 
U^\beta -\case{1}{2}Tg_{\alpha \beta }U^\alpha U^\beta } \right)},
\]
where  $T$ is the contraction of   $T_{\alpha \beta }$ and $U^\alpha$ is the four-velocity. 
Then for a gas with density  $\rho $ where the pressures are negligible the energy 
loss rate is (Crawford 1987a)
\[
\varepsilon c =-{{1} \over E}{{dE} \over {dt}}=\sqrt {{{8\pi G\left( {\rho 
c^2+p} \right)} \over {c^2}}},
\]
where x is measured along the photon's trajectory.  
This equation can be integrated to obtain
\[
E=E_0\exp (-\varepsilon \mathop x\limits_{}).
\]
If   $\rho =n\,m_H$  and with 
(using equation \ref{e3}) 
\[
p\approx p_c=-{{\rho c^2} \over {3}}\left( {\overline {\gamma ^2}-1} \right)=-{{1}
\over {3}}\rho c^2,
\]
then   $\varepsilon =4.54\times 10^{-27}\sqrt{n}\,\mbox{m}^{-1}$  and the predicted 
Hubble's  constant is 
\begin{equation}
\label{e24}
H=c\varepsilon =42.0\,\sqrt{n} \,\mbox{km}\, \mbox{s}^{-1}\, \mbox{Mpc}^{-1}.
\end{equation}
With the value   $n=2.05m_H \mbox{ m}^{-3}$  we get  $H=60.2\,\mbox{km}\,
\mbox{s}^{-1}\, \mbox{Mpc}^{-1}$.  Note for non-cosmological applications where 
the curvature pressure is negligible the results are   
$\varepsilon =5.57\times 10^{-27}\sqrt{n} \,\mbox{m}^{-1}$ or  
\begin{equation}
\label{e24a}
\varepsilon c = 51.5\sqrt{n} \,\mbox{km}\, \mbox{s}^{-1}\, \mbox{Mpc}^{-1}.
\end{equation}
Required later is the product of Hubble's constant with the radius of the universe 
which is  $RH=\sqrt{2}\,c$. This is identical to that derived earlier (Crawford 1993) 
for a Newtonian cosmology.

The principle of the  focusing theorem can be illustrated by considering a very long 
cylinder of gas and Newtonian gravitation. At the edge of the cylinder of radius  $r$ 
the acceleration to-wards the center of the cylinder is   $\ddot r=2\pi G\rho r$ 
where the dots denote differentiation with respect to time.  
Hence  for the area  A we get   $\ddot A=4\pi G\rho A$. 
Except for the numerical constant  this is the same as that for general relativity 
showing that it is the local density that determines focusing. The difference of a 
factor of one half is because the model only includes space curvature and not 
spacetime curvature. In both cases distant masses have no effect. 
In particular there is no focusing and hence no energy loss in the exterior 
Schwartzschild field of a spherical mass distribution such as the sun.

Since the excitation of the photon is slowly built up along its trajectory before  
the emission of two low energy photons any other interaction that occurs with a path 
length shorter than that between the emission of secondaries will clearly diminish 
their production. 
That is the excitation  can be dissipated without any extra energy loss. 
The average distance between emission of secondaries is (Crawford 1987a; using 
Heisenberg's uncertainty principle)  
$\Delta x=\sqrt {\lambda _0/4\pi \varepsilon }$ where $h$  Plank's constant, 
$\varepsilon $ is the fractional rate of energy loss per unit distance defined above and 
$\lambda _0$ is the wavelength of the primary photon.

The classic experiment of Pound and Snyder \cite{Pound65} is an example of how the hypothesis 
of a gravitational interaction  may be tested. 
They used the Mossbauer effect to measure the energy of 14.4\,KeV (${}^{57}$Co)  
gamma rays after they had passed up or down a 22.5\,m path in helium.  
Their result for the gravitational  redshift was in excellent agreement with the predicted  
fractional change in energy of  $-2.5\times 10^{15}$. 
The gravitational interaction theory predicts a fractional change in 
energy due to the gravitational interaction based on the density of  helium in the tube of   
$-1.25\times 10^{-12}$  which is considerably larger.  
Since their measurement was for the difference between upward and downward paths any 
effects independent of direction will cancel. 
However for these conditions although the typical path length between the emission of 
secondaries of 11m is less than the length of the apparatus it is still much longer 
than the mean free path for coherent forward scattering that is the quantum description 
of refractive index.  
In this scattering the  photon is absorbed by many electrons and after a short time 
delay (half a period) a new photon with the same energy and momentum is emitted. 
For these high energy gamma rays the binding energy of the electrons can be ignored 
and the mean free path for coherent forward scattering is given by the Ewald and Oseen 
extinction length \cite{Jackson75} of   $X=1/\left( {\lambda r_0n_e} \right)$ where   
$\lambda $ is the wavelength and   $r_0$ is the classical electron radius. 
In this case   $X=0.15\,$m that is much less than the 11\,m required for secondary 
emission and therefore the gravitational interaction energy loss will be minimal.  
The major difficulty with a laboratory test is 
in devising an experiment where   $\Delta x$ is less than the size of the apparatus 
and also less than the mean free path of any other interaction. 
Nevertheless if there are any residual effects they may be detectable  in such an 
experiment with a horizontal run using gases of different types and densities.

This inhibition of the gravitational interaction can occur in astrophysical situations. 
Consider the propagation of radiation through the galaxy where there is a fully 
ionized plasma with density  $\rho =n\,m_H$, then the critical density is defined by 
when the Ewald and Oseen extinction length is equal to the distance between emission 
of secondary photons. 
If the density is greater than this critical density then the inhibition by refractive 
index impairs the gravitational interaction and there is a greatly reduced redshift. 
The critical density (for a hydrogen plasma) is   $n_e=426.5/\lambda ^2\,\mbox{m}^{-3}$. 
For 21\,cm radiation the critical density is  $n_e=9,700\,\mbox{m}^{-3}$  and since most 
inter-stellar densities are much larger than this we do not expect 21\,cm radiation 
within the galaxy to show redshifts due to the gravitational interaction. Thus all 
redshifts of 21\,cm radiation within the galaxy are primarily due to doppler shifts.  
However optical radiation in the Galaxy should show the redshift due to the 
gravitational interaction. 

It has been argued \cite{Zel'dovich63} that tired light cosmologies (such as this) 
should show a smearing out of the images of distant sources. 
The argument is that if the energy loss is caused by an interaction with 
inter-galactic matter, it is accompanied by a transfer of momentum with a 
corresponding  change in direction. 

That is the photon is subject to multiple scattering and hence photons from 
the same source will eventually have slightly different directions and its image 
will be smeared. 
For the gravitational interaction  the interaction is not with some particle with 
commensurate mass but with the mass of the gas averaged over a suitable volume. 
Since the effective mass is so large the scattering angles will be negligible. 
Furthermore in low density gas the photon loses energy to two secondary photons 
and to conserve spin and momentum these must be emitted symmetrically so that 
there is no scattering of the primary photon.

\section{The microwave background radiation}
Because of their wave nature electrons and other particles will be subject to the 
focusing 
theorem in a way similar to photons. 
In Crawford \cite{Crawford91} it was argued that  particles such as electrons are subject to 
a similar centripetal acceleration that produces a fractional energy loss rate of 
$\varepsilon _e$, and for a gas with density  $\rho $ and pressure $p$ it is 
\[
\varepsilon _e=\sqrt {{{8\pi G} \over {c^2}}\left[ {(\gamma ^2 - \case{1}{2})\rho 
c^2+\left( {\gamma ^2 + \case{1}{2} } \right)p} \right]}\ ,
\]
where  $\gamma $ is the usual velocity factor. 
Hence the rate of energy loss as a function of distance  is
\[
{{dP^0} \over {dx}}=\sqrt {{{8\pi G} \over {c^4}}\left[ {(\gamma^2-
\case{1}{2})\rho c^{^2}+\left( {\gamma ^2+\case{1}{2}} \right)p} \right]}\,\beta^2P^0,
\]
where  $\beta =v/c$ is the particle's velocity relative to the medium and $P^0$ is 
the energy component of its momentum four-vector. 
As it moves along its trajectory the particle will be excited  by the focusing of 
its wave packet.  
For charged particles conservation of spin prevents them from removing their 
excitation by direct emission of low energy photons. 
However if there is a photon field it may interact with a photon in a process 
like stimulated 
emission and thereby lose energy.  
The dominant photon field in inter-galactic space is that 
associated with the microwave background radiation. 
The model proposed is that the electrons lose energy by stimulated emission  to 
the background radiation so that the local black body spectrum is conserved. 
Concurrently because of the gravitational interaction the photons are losing energy 
that is absorbed by the plasma. 
Note that most of the secondary photons have frequencies below the plasma frequency. 
Although this means that they cannot propagate it does not prevent direct absorption 
of their energy. 
After all for frequencies below the plasma frequency the electrons can have bulk 
motion and absorb energy from an oscillating field. 
Given an equilibrium condition in which the energy lost by the electrons is equal 
to the energy lost by the photons we can equate the two energy loss rates  and get 
an expression for the temperature of the microwave background radiation \cite{Crawford91} 
of
\[
T_M^4={{n_em_ec^3} \over {4\sigma }}\overline{ {\left( {(\gamma ^2-
\case{1}{2})+\left( {\gamma ^2+\case{1}{2}} \right){{p} \over {\rho c^{^2}}}} 
\right)\beta ^3\gamma } },
\]
where  $n_e$ is the electron number density,  $m_e$ is the electron mass,  
$\sigma $ is the Stefan-Boltzmann constant and, an average s done over all electron 
velocities.  
For an electron temperature of $1.11\times 10^9$\,K the bracketed term has the value 
of 0.555 with zero pressure or 0.412 with the gravitational curvature pressure.
With an electron density of 1.78\,m$^{-3}$ corresponding to a mass density of
$2.05m_H \mbox{ m}^{-3}$ and with the curvature pressure included  the predicted 
temperatures is  3.0\,K. 
Given the deficiencies of the model (mainly its assumption of homogeneity) this 
is in good agreement with the observed value of 2.726\,K \cite{Mather94}. 
It is interesting that the predicted temperature only depends on the average 
density and a function of electron velocities that 
is of order one.

\section{No dark matter}
In the standard big-bang cosmology there are three major arguments \cite{Trimble87} for the 
existence of dark matter, that is matter that has gravitational importance  but is 
not seen at any wavelength. 
The first argument is based on theoretical considerations of closure and reasonable 
cosmological models within the big-bang paradigm.  
The second is from the application of the virial theorem to clusters of galaxies and 
the third is that galactic rotation curves show high velocities at large radii. 
The first of these is purely an artifact of the big-bang cosmological model; it is not 
based on observation  and therefore it is not relevant to this cosmology. 
The second and third are based on observations and will be discussed at some length.

In the standard big-bang model all the galaxies in a cluster are gravitationally bound 
and do not partake in the universal expansion. 
If they are gravitationally bound then assuming that their differential (peculiar) 
redshifts are due to differential velocities we can use the virial theorem to estimate 
the total (gravitational) mass in the cluster. 
Typically this gravitational mass is one to several orders of magnitude larger than 
the mass derived from the luminosities of the galaxies: hence the need for dark matter. 

Observations of X-rays from galactic clusters show that there is a large mass of gas in the 
space between the galaxies.  
Although the mass of this inter-cluster gas is small compared to the mass of the presumed 
dark matter it is large enough to give significant redshifts due to the gravitational 
interaction. 
Thus the current model ascribes most of the differential redshifts to gravitational 
interactions in the inter-cluster gas. 
This model has been quantitatively investigated by Crawford \cite{Crawford91}
for the Coma cluster. 
The method used was to take the observed differential redshift for each galaxy and by 
integrating equation (\ref{e24a}) through the known inter-galaxy gas the differential 
line-of-sight distance to the galaxy was computed. 
The gas density distribution that was used is that given by Gorenstein, 
Huchra \& de Lapparent \cite{Gorenstein79}. 
The result  is that galaxies with lower redshifts than that for the center of the 
cluster would be nearer and those with higher redshifts would be further away. 
The model assumed that the inter-cluster gas was spherically distributed and the 
test was in how well the distribution of Z coordinates  compared with those for 
the X and Y coordinates that were in the plane of the sky.   
Furthermore it was assumed that genuine velocities were negligible compared to 
the effective velocities of the differential redshifts. 
The median distances for each coordinate were X=0.19\,Mpc, Y=0.17\,Mpc and Z= 0.28\,Mpc. 
Given that the Coma cluster has non-spherical structure and that the model is 
very simple the agreement of the median Z distance with those for X and Y is good. 
Again it should be emphasized that there were no free parameters; the Z distances 
depend only on the gas distribution, the measured differential redshift, and 
equation (\ref{e24a}). 
If this result can be taken as representative of clusters  then there is no need 
for dark matter to explain cluster `dynamics'. 
The large differential redshifts are mainly due to gravitational interactions in 
the inter-galactic gas.

One of the difficulties with the big-bang cosmology is that it is so vague in its 
predictions that it is very difficult to refute it with observational evidence. 
However the redshifts from a cluster of galaxies can provide a critical test. 
Since celestial dynamics is time reversible a galaxy at any point in the cluster is 
equally likely to have a line-of sight velocity towards us as away from us.  
Then if accurate measurements of magnitude, size or some other variable can be used 
to get differential distances there should (in the big-bang cosmology) be no correlation 
between differential redshift and distance within the cluster.
Whereas in the static cosmology proposed here there should be a strong correlation 
with the more distant galaxies having a higher differential redshift. 
Clearly this is a difficult experiment since for the Coma cluster  it requires 
measurements of differential distances  to an accuracy of about 1\,Mpc at a 
distance of 100\,Mpc.

The third argument for dark matter comes from  galactic rotation curves. 
What is observed is that  velocity plotted as a function of distance along the 
major axis shows the expected rapid rise from the center but instead of reaching 
a maximum and then declining in an approximately Keplerian manner it tends to stay 
near its maximum value. 
The standard explanation is that there is a halo of dark matter that extends well 
beyond the galaxy and that has a larger mass than the visible galaxy. 
For this static cosmology a partial explanation is that most of the redshift is 
due to gravitational interaction in a halo but one that is commensurate in size 
with the galaxy. 
Although it is possible to devise density distributions that can explain particular 
rotation curves there is no universal model that can explain all rotation curves. 
The conclusion is that there is a mechanism that could explain some rotation curves 
but whether it can explain all rotation curves remains to be seen.

These two cases illustrate an important aspect of redshifts in this cosmology. 
Although the 
redshift is on average an excellent measure of distance any particular redshift is 
only a measure of the gas in its line of sight. 
Any lumpiness in the inter-cluster gas will produce apparent structure in redshifts 
that could be falsely interpreted as structure in galaxy distributions. 
That is, the apparent "walls", "holes", and other structures may be due to intervening 
higher density  or lower density clouds. 
For example the model predicts an apparent hole behind clusters of galaxies because of 
gravitational interactions in  intra-cluster gas. 
The velocity width of the hole would be of the same magnitude as the velocity dispersion 
in the cluster. 
For the Coma cluster the velocity width of this hole would vary from about 
4100\,kms$^{-1}$ near the center of the cluster to about 1200\,kms$^{-1}$ near the edge. 

\section{No evolution}
The most important observational difference between this cosmology and the big-bang 
cosmology is that it obeys the perfect cosmological principle: it is homogeneous both 
in space and time. 
Consequently any unequivocal evidence of evolution would be fatal to its viability.  
In contrast the big-bang theory demands evolution. 
However it has the  difficulty that the theory only provides broad guides as to what 
that evolution should be and there is little communality between the  evolution 
required for different observations.  
Nevertheless there is an entrenched view that evolution is observed in the 
characteristics of many objects. 
Two notable examples are the luminosity distribution of quasars and the angular-size 
relationship for radio galaxies. 
It will be shown that the observations for both of these phenomena are fully compatible 
with a static cosmological model. 

\section{Quasar luminosity distribution}
Because of their high redshifts quasars are excellent objects for probing the distant universe. 
Since this cosmological model is static neither the density distribution nor the luminosity 
distribution of any object should be a function of distance. 
Consider the density distribution  
$n\left(z\right)$ where z is the usual redshift parameter 
 $z=(\lambda _{\mbox{observed}}/\lambda _{\mbox{emitted}}-1$) then 
\[
z=\exp \left( {Hr/c} \right)-1,
\]
where $r$ is  the distance. 
Since the range of $r$ is  $0\le r\le \pi R$ the maximum value of z is 84.0 and its 
value at the `equator' is  8.2. 
Given that the geometry is that for a 
three-dimensional hyper-spherical surface with radius R in a four-dimensional space 
the volume out to a distance  $r$ is
\[
V\left( r \right)=2\pi R^2\left( {r-{{R} \over {2}}\sin \left( {{{2r} \over {R}}} 
\right)} \right)
\]
and the density distribution as a function of redshift for an object with a uniform 
density of   $n_0$ 
\begin{eqnarray}
\label{e30}
n\left( z \right)dz&=&n_0{{dV} \over {dr}}{{dr} \over {dz}}dz\nonumber \\
&=&{{4\pi R^2cn_0\sin ^2\left( {c\ln \left( {1+z} \right)/RH} \right)} \over {H\left( {1+z} 
\right)}}dz.
\end{eqnarray}
From equations (\ref{e14}) and (\ref{e24}) we find that  $HR=\sqrt 2\,c$ and 
equation (\ref{e30}) becomes 
 \begin{equation}
\label{e31}
n\left( z \right)dz={{4\pi R^3n_0\sin ^2\left( {\ln \left( {1+z} \right)/\sqrt 2} 
\right)} \over {\sqrt 2\left( {1+z} \right)}}dz,
\end{equation}
which has a maximum when  $z=2.861$. 
Now the difficulty of using equation (\ref{e31}) with observations is that most 
quasar observations  have severe selection effects. 
Boyle et al \cite{Boyle90} measured the spectra of 1400 objects of which 351 were identified 
as quasars with redshifts  z $<$ 2.2. 
The advantage of their observations is that their selection effects were well defined. 
A full analysis is given by Crawford \cite{Crawford95b}. 

Let a source have a luminosity  $L\left( \nu \right)$(Whz$^{-1}$) at the emission 
frequency $\nu$. 
Then if the energy is conserved the observed flux density 
$S\left( \nu  \right)$ (Wm$^{-2}$Hz$^{-1}$) at a distance  $r$  is the luminosity 
divided by the area which is
\[
S\left( \nu  \right)={{L\left( \nu  \right)} \over {4\pi R^2\sin ^2\left( {r/R} 
\right)}}.
\]
However because of the gravitational interaction there is an energy loss such that 
the received frequency   $\nu _0$ is related to the emitted frequency   $\nu _e$ by 
\[
\nu _0=\nu _e\exp \left( {-Hr/c} \right)=\nu _e/\left( {1+z} \right).
\]
This  loss in energy means that the observed flux density is decreased by a factor of  $1+z$. 
But there is an additional bandwidth factor  that exactly balances the energy loss
factor. 
In addition allowance must be made for K-correction \cite{Rowan-Robinson85} that relates 
the observed spectrum to the emitted spectrum. 
Since it is usual to include the bandwidth factor in the K-correction the apparent magnitude is
 \begin{eqnarray*}
m & =&  -\case{5}{2}\log \left( {S\left( {\nu _0} \right)} \right)\nonumber \\
  & =&  -\case{5}{2}\log \left( {L\left( {\nu _0} \right)} \right)+
\case{5}{2}\log \left( {4\pi R^2} \right)
\nonumber \\
&+&
5\log \left( {\sin \left( {{{c\ln (1+z)} 
\over {HR}}} \right)} \right) \\
&+&\case{5}{2}\log \left( {1+z} \right)
+K\left( z \right),
\end{eqnarray*}
where  $K\left( z \right)$ is the K-correction. 
The result of the analysis was that the observations were fitted by a (differential)  
luminosity function  that had a Gaussian shape with a standard deviation of 1.52 magnitudes 
and a maximum at  $M=-22.2\,\mbox{mag}$ (blue). 
The only caveat was that there appeared to be a deficiency of weak nearby quasars in the sample. 
Since all cosmological models are locally Euclidean this must be a selection effect.
The fact that the absolute magnitude distribution had a well-defined peak and this was 
achieved without requiring any  evolution is strong support for the static model.

\section{Angular size of radio sources}
For the geometry of the hyper-sphere the observed angular size   $\theta $ for an object 
with a redshift of  z and projected linear size of D is   
$\theta =D/\left( {R\sin \left( {r/R} \right)} \right)$, and in terms of redshift it is 
 \begin{eqnarray*}
\theta &=&{D \over {R\sin \left( {c\ln \left( {1+z} \right)/RH} \right)}}\\
&=&{D \over {R\sin \left( {\ln \left( {1+z} \right)/\sqrt 2} \right)}}.
\end{eqnarray*}
The angular size decreases with z until   $z=8.2$ where there is a broad minimum and 
then it increases again. 
This model was used by Crawford \cite{Crawford95a} to analyze 540 double radio sources 
(all Faranoff-Riley type II) listed by Nilsson et al \cite{Nilsson93}. 
The result was an excellent fit to the radio-source size measurements, much 
better than the big-bang model with a free choice of its acceleration parameter.  

\section{Other evidence for evolution}
There are however more direct observations of evolution that will be discussed. 
They are the 
distribution of absorption lines in quasar spectra, the measurement of the 
microwave background temperature at high redshift, and the time dilation of the 
type I supernova light curves at large distances. 
For this static cosmology consider a uniform distribution of objects with number 
density N and cross-sectional area A  then their distribution in redshift along
 a line of sight is (here   $\gamma $ is the exponent and not the Lorentz velocity parameter)
\[
{{dN} \over {dz}}={{NAc} \over {H}}\left( {1+z} \right)^\gamma \ .
\]
with $\gamma =-1$. 
If the  absorption lines seen in the spectra of quasars are due to absorption by 
the Lyman-$\alpha$ line of hydrogen in intervening clouds of gas and with a 
uniform distribution of clouds their predicted redshift  distribution should 
have $\gamma =-1$. 
However observations 
\cite{Hunstead88,Morris91,Williger94,Storrie-Lombardi97} 
show exponents that range from 0.8 to 4.6. 
Although there is poor agreement amongst the observations clearly they are all in 
disagreement with this model. 
Observations of absorption lines have complications due to lack of resolution causing 
lines to be merged and that only a limited range in z (from Lyman-$\alpha$ to Lyman-$\beta$)  
is available from each quasar. 
However the major change required in the interpretation of the results  for the static 
cosmology  is in the explanation for the broad absorption lines. 
Traditionally these have been ascribed to Doppler 
broadening from bulk motions in the clouds but it is also possible that they are due 
to energy loss by the gravitational interaction. 
For example using equation (\ref{e24a}) the `velocity' width of a cloud of 
diameter $10^4$\,pc and density  $10\,m_Hm^{-3}$ is  $16\,\mbox{km}\, \mbox{s}^{-1}$ 
which is typical of the observed line widths. 
For a typical column density of  $ N_{\mbox{\ion{H}{1}}}=10^{15}\,\mbox{cm}^{-2}$ 
this cloud would have a  ratio of \ion{H}{1} to ionized hydrogen of  
$3\times 10^{-5}$. 
A further consequence is that because of the clouds the observed redshift 
is not a valid measure of the true distance. 
For example suppose the quasar is located in a galactic cluster where we would expect 
a high local concentration of clouds then its redshift would be increased over that 
expected for the cluster by the extra energy loss in the clouds. 
The conclusion is that until the nature if the absorption lines are better understood 
and analyzed in the context of this theory the evidence for evolution is not convincing.

Another observation that could refute this theory is if the cosmic microwave background 
radiation has a higher temperature at large distances. 
Ge et al \cite{Ge97} measured the absorption from the ground and excited states of
C1 (with a redshift of 1.9731)  in the quasar  QSO 0013-004. 
They measure the strengths of the J=0 and J=1 fine structure levels  and derived an 
excitation temperature of   $11.6\pm 1.0$\,K which after corrections gives a temperature 
for the surrounding radiation of  $7.9\pm 1.0$\,K that is in good agreement with the 
redshifted temperature of 8.1\,K. 
On face value this is clear evidence for evolution. 
But not only are the measurements difficult they are based on a model for line widths 
that does not include the gravitational interaction. 
Until this is done and the results are confirmed for other quasars and by other 
observers a static cosmology is not refuted. 

Programs that search for supernovae in high redshift galaxies with large telescopes are now 
finding many examples and more importantly some are being detected before they reach 
their maximum intensity. 
Leibundgut et al \cite{Leibundgut96},  Goldhaber et al \cite{Goldhaber96}, 
and Riess et al \cite{Riess97} have reported 
on type 1a supernovae in which they believe that they can identify the type of supernova 
from its spectral response and by comparing the supernova light curves with reference 
templates they measure a time dilation that corresponds to that expected for their 
redshift in a big-bang cosmology.  
However because of uncertainties in matching the exact type of supernova and because 
of the occurrence of individual inhomogeneities many more observations are needed 
before these results are well established.

The conclusion is that the Lyman-$\alpha$ forest observations and the cosmic background 
radiation 
temperature observations need to be re-evaluated within  the static cosmological model 
in order to see if they show evolution and refute the model. 
The supernovae results are essentially unchanged in the static model and if they hold 
up they make a strong case for evolution that would refute any static model.

\section{Nuclear abundance}
In this cosmology the universe is dominated by a very high temperature plasma. 
Galaxies 
condense from this plasma, evolve and die. 
Eventually all of their matter is returned to the plasma.  
Although nuclear synthesis in stars and supernova can produce the heavy elements it 
cannot produce the very  light elements. 
In big-bang cosmology these are produced  early in the expansion when there were high 
temperatures and a large number of free neutrons. 
This mechanism is not available in a static cosmology. 
Nevertheless the temperature of the plasma  ($2\times 10^9\,\mbox{K}$)  is high enough 
to sustain nuclear reactions. 
The problem is that the density is so low that reaction rates will be minuscule. 
One important reaction is the photo-disintegration of heavy nuclei by the background 
X-ray radiation. 
As heavy nuclei are returned to the plasma they are broken down by interactions with 
the radiation to produce lighter nuclei. 
The end result is an abundance distribution dominated by hydrogen and with smaller 
quantities of helium and other light elements. 
Naturally much further work is needed to quantify this hypothesis.

\section{Entropy}
Nearly every textbook  on elementary physics quotes a proof based on  the second law of 
thermodynamics to show that the entropy of the universe is increasing but this is in 
direct conflict with the perfect cosmological principle  where total entropy is constant. 
The conflict can be resolved if it is noted that the formal proof of the second law of 
thermodynamics requires consideration of an isolated system and the changes that occur 
with reversible and irreversible heat flows between it and its surroundings. 
Now there is no doubt that irreversible heat flows occur and lead to an overall 
increase in entropy. 
However the formal proof is flawed in that with gravitational fields one cannot 
have an isolated system. 
There is no way to shield gravity. 
Furthermore in their delightful  book Fang \& Li \cite{Fang89} argue that a self-gravitating 
system has negative thermal capacity and that such systems cannot be in thermal equilibrium.  
The crux of their argument is that if energy is added to a self-gravitating system, 
such as the solar system, then the velocities and hence the `temperature' of the bodies decrease. 
What happens is that from the virial theorem the potential energy (with a zero value 
for a fully dispersed system)  is equal to minus twice the kinetic energy and the 
total energy is the sum of the potential and kinetic energies which is therefore 
equal to minus the kinetic energy. 
Thus we must be very careful in applying simple thermodynamic laws to gravitational systems. 

Now consider the gravitational interaction  where photons lose energy to the background 
plasma. 
Since this process does not depend on temperature it is not a flow of heat energy rather 
it is work. 
Nevertheless we can compute the entropy loss from the radiation field, considered as a 
heat reservoir, as  $-W/T_r$ where W is the energy lost, and similarly the entropy 
gained by the plasma as  $W/T_e$. 
Then since  $T_e>>T_R$ there is a net entropy loss. 
Thus this gravitational interaction not only produces the Hubble redshift but it also 
acts to decrease the entropy of the universe thereby balancing the entropy gained in 
irreversible processes such as the complementary interaction where electrons lose 
energy to the radiation field.  

\section{Olber's paradox}
An essential requirement of any cosmology is to be able to explain Olber's paradox (or more 
correctly de Chesaux's paradox \cite{Harrison81}) as to why the sky is dark at night. 
For the big-bang cosmology although the paradox is partly explained by the universal 
redshift the major  reason is that the universe has a finite lifetime. 
For this static cosmology the explanation is entirely due to the redshift. 
The further we look  to distant objects the more the light is redshifted until it is 
shifted outside our spectral window. 
Thus in effect we only see light from a finite region.  
Note that the energy lost by the photons is returned to the inter-galactic plasma as 
part of a cyclic process. 

\section{Conclusion}
The introduction of curvature pressure has wide ranging astrophysical applications. 
It is possible that it may resolve the solar neutrino problem but this must await a full 
analysis using the standard solar model.  
Although the theory does not prevent the formation of a black hole from cold matter 
it does have an important effect on the high temperature accretion rings and may provide 
the engine that produces astrophysical  jets. 

The greatest strength of this model is that it shows how a stable and static cosmology 
may exist within the framework of general relativity without a cosmological constant.  
The model with a homogeneous plasma depends only on one parameter, the average density 
which from X-ray observations is taken to be  $2.05m_H \mbox{ m}^{-3}$. 
It then predicts that the plasma has a temperature of  $2\times 10^9\,K$ and that the 
universe has a radius given by equation (\ref{e14}).  
It has been shown that for a simple inhomogeneous density distribution the predicted 
temperature could easily be much lower and it could be in agreement with the temperature 
observed for the X-ray background radiation.  
Inclusion of the gravitational interactions permits the prediction of a Hubble constant 
of $H=60.2\,\mbox{km}\, \mbox{s}^{-1}\, \mbox{Mpc}^{-1}$ and a microwave background 
radiation with a temperature of  $3.0\,K$.  
Dark matter does not exist but arises from assuming that non-cosmological redshifts are 
genuine velocities and then using  the virial theorem. 
In this static model most of the non-cosmological  velocities are due to gravitational 
interactions in intervening  clouds. 

Analysis of the observations for quasar luminosities and  the angular size of radio sources 
shows that they can be fully explained in a static cosmology without requiring any evolution. 
The implication is that many other observations that require evolution in the big-bang 
cosmology need to be re-examined within the static cosmology before evolution can be confirmed.  
The strong evolution shown in the distribution of absorption lines (the Lyman-$\alpha$ forest) 
is a problem for the static model. 
However because of the gravitational interaction that can cause line broadening and the
possibility that some of the redshift may come from the clouds that produce the absorption 
lines the results cannot at this stage be taken as a refutation of the static model.
Although the observations of a redshifted background microwave temperature and the 
evidence of time dilation in the decay curves of type 1a supernovae appear to show 
direct evolution  it is too early to be certain.  
These observations need  better statistics and should be analyzed within this static 
model before their apparent evolution is convincing. 

The model includes a qualitative model for the generation of the light elements in the high 
temperature inter-galactic plasma. 
The most important interaction is probably the photo-disintegration of heavy elements 
using the background X-ray radiation.   
It was also argued that the effects of gravitational interaction of the microwave 
background radiation that transfers energy to the high temperature plasma decreases 
entropy so that overall total entropy of the universe is constant. 
Finally the sky is dark at night because the light from distant stars is redshifted 
out of our spectral window.  

An important characteristic of this static cosmology is that it is easily refuted: 
any unequivocal evidence for evolution would disprove the model. 
Apart from evolution the most discriminating test that chooses between it and the 
big-bang cosmology would be to compare the differential velocities of galaxies in a 
cluster with their distance. 
Whereas the big-bang model  requires that there is no correlation this static cosmology 
requires that the more distant  galaxies will have larger redshifts.

\section{Acknowledgments}
This work is supported by the Science Foundation for Physics within the 
University of Sydney, and use has made of NASA's Astrophysics Data System Abstract Service. 


\begin{thebibliography}{10}

\bibitem{Crawford87a}
D.~F. Crawford, Aust. J. Phys. {\bf 40},  449  (1987).

\bibitem{Crawford91}
D.~F. Crawford, Astrophys. J. {\bf 377},  1  (1991).

\bibitem{Crawford93}
D.~F. Crawford, Astrophys. J. {\bf 410},  488  (1993).

\bibitem{Misner73}
C.~W. Misner, K.~S. Thorne, and J.~A. Wheeler, {\em Gravitation} (W. H. Freeman
  and Co., San Francisco, 1973).

\bibitem{Groot80}
S.~R. de~Groot, W.~A. Leeuwen, and C.~G. van Weert, {\em Relativistic Kinetic
  Theory} (North-Holland Pub. Co., Amsterdam, 1980).

\bibitem{Abramowitz72}
M. Abramowitz and I.~A. Stegun, {\em Handbook of Mathematical Functions}, 2nd
  ed. (Dover Publications, N. Y., 1972).

\bibitem{Bahcall89}
J. Bahcall, {\em Neutrino Astrophysics} (Cambridge University Press, Cambridge,
  1989).

\bibitem{Bahcall97}
J. Bahcall,  in {\em Proc. of the 18th Texas Symposium on Relativistic
  Astrophysics} (World Scientific, Singapore, 1997).

\bibitem{Rees97}
M.~J. Rees,  in {\em Proc. of Chandrasekhar Memorial Conf.} (Chicago, Chicago,
  1997).

\bibitem{Begelman96}
M. Begelman and M.~J. Rees, {\em Gravity's fatal attraction. Black holes in the
  universe} (W. H. Freeman, New York, 1996).

\bibitem{Kormendy95}
J. Kormendy and D. Richstone, Ann. Rev. Astron. Astrophys. {\bf 33},  581
  (1995).

\bibitem{Weinberg72}
S. Weinberg, {\em Gravitation and Cosmology} (John Wiley and Sons, New York,
  1972).

\bibitem{Fabian92}
A.~C. Fabian and X. Barcons, Ann. Rev. Astron. Astrophys. {\bf 30},  429
  (1992).

\bibitem{Crawford87b}
D.~F. Crawford, Aust. J. Phys. {\bf 40},  459  (1987).

\bibitem{Allen76}
C.~W. Allen, {\em Astrophysical Quantities}, 3rd  ed. (The Althone Press,
  London, 1976).

\bibitem{Mather94}
J.~C. Mather {\it et~al.}, Astrophys. J. {\bf 420},  439  (1994).

\bibitem{Peebles93}
P.~J.~E. Peebles, {\em Principles of Physical Cosmology} (Princeton University
  Press, Princeton, New Jersey, 1993).

\bibitem{Raychaudhuri55}
A.~K. Raychaudhuri, Phys. Rev. {\bf 98},  1123  (1955).

\bibitem{Pound65}
R.~V. Pound and J.~L. Snider, Phys. Rev. B {\bf 140},  788  (1965).

\bibitem{Jackson75}
J.~D. Jackson, {\em Classical Electrodynamics} (John Wiley, New York, 1975).

\bibitem{Zel'dovich63}
Y.~B. Zel'dovich, JETP {\bf 16},  1395  (1963).

\bibitem{Trimble87}
V. Trimble, Annu. Rev. Astron. Astroph. {\bf 25},  425  (1987).

\bibitem{Gorenstein79}
P. Gorenstein, D. Fabricant, K. Topka, and F.~R. {Harden, Jr.}, Astrophys. J.
  {\bf 230},  26  (1979).

\bibitem{Boyle90}
B.~J. Boyle, R. Fong, T. Shanks, and B.~A. Peterson, Mon. Not. R. Astron. Soc.
  {\bf 243},  1  (1990).

\bibitem{Crawford95b}
D.~F. Crawford, Astrophys. J. {\bf 441},  488  (1995).

\bibitem{Rowan-Robinson85}
M. Rowan-Robinson, {\em The Cosmological Distance Ladder} (W. H. Freeman and
  Co., New York, 1985).

\bibitem{Crawford95a}
D.~F. Crawford, Astrophys. J. {\bf 440},  466  (1995).

\bibitem{Nilsson93}
K. Nilsson, M.~J. Valtonen, J. K0tilainen, and T. Jaakkola, Astrophys. J. {\bf
  413},  453  (1993).

\bibitem{Hunstead88}
R.~W. Hunstead, H.~S. Murdoch, M. Pettini, and J.~C. Blades, Astrophys. J. {\bf
  329},  527  (1988).

\bibitem{Morris91}
S.~L. Morris, R.~J. Weymann, B.~D. Savage, and R.~L. Gilliland, Astrophys. J.
  {\bf 377},  L21  (1991).

\bibitem{Williger94}
G.~M. Williger {\it et~al.}, Astrophys. J. {\bf 428},  574  (1994).

\bibitem{Storrie-Lombardi97}
L.~J. Storrie-Lombardi, R.~G. McMahon, M.~J. Irwin, and C. Hazard,  in {\em ESO
  Workshop on QSO Absorption Lines} (in press, ADDRESS, 1997).

\bibitem{Ge97}
J. Ge, J. Bechtold, and J.~H. Black, Astrophys. J. {\bf 474},  67  (1997).

\bibitem{Leibundgut96}
B. Leibundgut {\it et~al.}, Astrophys. J. {\bf 466},  L21  (1996).

\bibitem{Goldhaber96}
G. Goldhaber {\it et~al.},  in {\em Thermonuclear Supernovae (NATO ASI)} (in
  press, ADDRESS, 1996).

\bibitem{Riess97}
A.~G. Riess {\it et~al.}, aj {\bf 114},  722  (1997).

\bibitem{Fang89}
L.-Z. Fang and S.~X. Li, {\em Creation of the Universe} (World Scientific,
  Singapore, 1989).

\bibitem{Harrison81}
E.~R. Harrison, {\em Cosmology} (Cambridge University Press, Cambridge, 1981).

\end{thebibliography}

\end{document}